\documentclass{iaus}
\usepackage{graphicx}

\def \aap{AAP}
\def \apjl{ApJ}
\def \apj{ApJ}

\def \araa{AAP}

\def \mnras{MNRAS.}

\def \nat{Nature}

\def\msun{{\,M_\odot}}

\title[~~The Galactic Centre as Starburst Laboratory] 
{The Galactic Centre - A Laboratory for Starburst Galaxies (?)}

\author[Roland M. Crocker]   
{Roland M. Crocker$^1$
}

\affiliation{$^1$Max-Planck-Institut f{\" u}r Kernphsik, P.O. Box 103980 \\Heidelberg, Germany\\
email: {\tt Roland.Crocker@mpi-hd.mpg.de} \\[\affilskip]
}

\pubyear{2011} 
\volume{284}  
\pagerange{1--12}
\jname{The Spectral Energy Distribution of Galaxies}
\editors{R.J. Tuffs \&  C.C.Popescu, eds.}
\begin{document}

\maketitle

\begin{abstract}
The Galactic centre -- as the closest galactic nucleus -- holds both intrinsic interest and possibly represents a useful analogue to star-burst nuclei which we can observe with orders of magnitude finer detail than these external systems. 
The environmental conditions in the GC -- here taken to mean the inner 200 pc in diameter of the Milky Way -- are extreme with respect to those typically encountered in the Galactic disk. 
The energy densities of the various GC ISM components are typically $\sim$two orders of magnitude larger than those found locally and the star-formation rate density $\sim$three orders of magnitude larger. 
Unusually within the Galaxy, the Galactic centre exhibits hard-spectrum, diffuse TeV (=$10^{12}$ eV) gamma-ray emission spatially coincident with the region's molecular gas. 
Recently the nuclei of local star-burst galaxies NGC 253 and M82 have also been detected in gamma-rays of such energies. 
We have embarked on an extended campaign of modelling the broadband (radio continuum to TeV gamma-ray), non- thermal signals received from the inner 200 pc of the Galaxy.
 On the basis of this modelling we find that star-formation and associated supernova activity is the ultimate driver of the region's non-thermal activity. 
 This activity drives a large-scale wind of hot plasma and cosmic rays out of the GC. 
 The wind advects the locally-accelerated cosmic rays quickly, before they can lose much energy in situ or penetrate into the densest molecular gas cores where star-formation occurs. 
 The cosmic rays can, however, heat/ionize the lower density/warm $H_2$ phase enveloping the cores. 
 On very large scales ($\sim$10 kpc) the non-thermal signature of the escaping GC cosmic rays has probably been detected recently as the spectacular `Fermi bubbles' and corresponding `WMAP haze'.
\end{abstract}

\firstsection 
\section{Introduction}

In part, this talk constitutes a general advertisement to the wider astronomical community concerning the utility of $\gamma$-ray data: analysis of GeV and TeV data -- especially combined with data at other wavelengths -- sometimes allows one to constrain the processes occurring or conditions prevailing in astrophysical regions with much greater precision than when recourse is made only to data in `traditional' wavebands.
I will illustrate this general philosophy through an analysis of the inner $\sim 200$ pc in diameter of our own Galaxy (hereafter the `Galactic centre' or GC).
To be sure, this region is important in its own right: as, by definition, the closest galactic nucleus our `up-close' view of the GC is of great interest 
for what it tells us -- by analogy -- about the activity of galactic nuclei in a cosmological context.
But an understanding of the GC -- responsible for $\sim$ 10\% of the Galaxy's {\it massive} star-formation (\cite{Figer2004}) and, of course, the host of its supermassive black hole -- 
is a prerequisite for understanding the overall ecology of the Galaxy as I hope to make clear below.

The ISM conditions in the GC are certainly extreme in comparison with those encountered in the Galactic disk: the energy densities/pressures of the various ISM phases are  two orders of magnitude higher than in the Galaxy-at-large, typically 100's of eV per cc.
The magnetic field amplitude, for instance, through the region is at least 50 $\mu$G (\cite{Crocker2010a}).
Equally, the plasma pressures and temperatures, light field, and gas turbulence are all much higher than locally (and all roughly in equipartition with each other).
All this is driven by star-formation which peaks sharply over the same inner $\sim$200 pc region of the Galaxy to a star-formation-rate areal density {\it three} orders of magnitude than the average value in the disk (e.g., \cite{Crocker2011b}).

Extending this general picture that the conditions in the GC may more closely resemble those encountered in a starburst environment, my recent work with collaborators (\cite{Crocker2010b,Crocker2011a,Crocker2011b})) has revealed a number of interesting facts about the GC:
i) the star-formation (and resultant) supernova activity in the GC drives a `super'-wind out of the region (more correctly, an outflow rather than a wind as the ejected material does not reach escape velocity);
ii) the GC outflow advects co-mingled plasma and non-thermal particle populations to large distance from the plane and -- I claim -- the non-thermal $\gamma$-ray and microwave
 signatures of this process have recently been found;
and iii)  despite the broad similarity to a star-burst alluded to above, the GC  is basically  in steady state: the current star-formation rate and consequent activity is typical for the time-averaged state of the GC over timescales approaching 10 Gyr.

\section{Multi-Wavelength Indications of a GC Outflow}

Unusually in the Galaxy, the GC is a source of extended, {\it diffuse} TeV emission as revealed by the HESS telescope (\cite{Aharonian2006}).
This emission, extending over $\sim 1.5^\circ$ in Galactic longitude, is spatially correlated with the column of molecular gas, showing peaks corresponding to the positions of the densest parts of the giant molecular complexes -- Sgr. B, C, etc -- inhabiting the region.
Such a correlation is expected in the case that the observed $\gamma$-rays originate from {\it hadronic} interactions, i.e., the collisions between non-thermal protons (and heavier ions in general) and ambient gas, exactly the collision processes typically investigated at the LHC.

On even wider scales than for the TeV emission ($\sim 6^\circ$ in Galactic longitude: \cite{LaRosa2005,Crocker2010a}),  radio continuum observations show that the GC is a distinct source of diffuse, $\sim$GHz, non-thermal emission.
Such emission must be due to the synchrotron losses experienced by a wide-spread population of cosmic ray electrons inhabiting the GC.

Despite the fact of this wide-spread non-thermal emission, the GC is actually significantly {\it under}luminous in both radio continuum and $\sim$TeV (and $\sim$GeV) $\gamma$-ray wavebands -- given the amount of star-formation currently going on there -- as we now explain.
Firstly, placing the GC on a plot of  its 60 $\mu$m vs. 1.4 GHz luminosity, one determines that radio continuum emission from this system falls one order of magnitude  (i.e., $\sim4 \sigma$) short with respect to the expectation afforded by the FIR-radio continuum correlation (e.g., \cite{Condon1992}).

Why might this be?
Although many details about the FIR-RC correlation remain unclear -- including why the correlation is, in general, so tight over so many orders of magnitude in luminosity (see contribution by T. Thompson to this volume) -- it must be that the correlation is ultimately related to the phenomenon of {\it massive} star-formation in galaxies.
Massive stars and their products dominate both the thermal and non-thermal radiative output from systems.
On the one hand, they
emit  most of the  optical and UV photons that are -- to a greater or lesser degree -- reprocessed by ambient dust into infrared wavelengths; FIR can, therefore, generally be treated as a direct tracer of the current star-formation rate in the system.
On the other, the lives of massive stars end in supernova explosions whose shock waves, expanding into the local ISM, are the sites where non-thermal particle populations (i.e., cosmic rays) are accelerated.
The subsequent synchrotron emission emitted by cosmic ray electrons as they gyrate around the local magnetic field dominates, in general, the $\sim$GHz radio continuum output of star-forming galaxies.
Given, then, that both sources of radiation are tied back to the rate at which massive stars are being formed, in the case that a system is calorimetric \cite{Voelk1989} to the electron population it accelerates (i.e., such electrons lose their energy radiatively in situ rather than carrying it away), there should be a correlation FIR and RC emission as observed.

Equally, given that cosmic ray hadrons are accelerated by the same supernova shocks, were galaxies or systems or calorimetric to these accelerated cosmic ray ion populations, then one would expect a linear scaling (\cite{Thompson2007}) between their (non-thermal) $\gamma$-ray  and their FIR emission.
Comparison between the GC system and expectation from this theoretical scaling places the GC's TeV emission at only $\sim 1$\% of expectation (with the system's GeV emission as detected by the $Fermi$ satellite (\cite{Chernyakova2010}) at about 10\% of expectation, but substantially polluted by point sources in the field)

What explains this phenomenology -- why is the GC so under-luminous in non-thermal emission?
Generically, three explanations present themselves:
\begin{enumerate}
\item The system suffered a significant a star-burst event more recently than the $10^7$ year lifetimes of the massive stars which dominate its radiative output. 
Of course, there is some stochasticity in the GC's star-formation rate: the GC has apparently undergone individual bursts of star-formation that have led to the creation of its super-stellar clusters (GC, Quintuplet, Arches) on these sort of timescales.
The evidence, however, from stellar luminosity function studies is that the rate of star-formation inferred from the creation timescale of these clusters is not atypical of (in fact somewhat less than) the GC's long-term averaged star-formation rate (\cite{Figer2004}).
In fact, the evidence that the GC's star-formation rate has been stable for a timescale approaching 10 Gyr (\cite{Figer2004,Maness2007}) seems increasingly firm.
So this explanation does not fly.

\item A second possibility is that conditions in the GC render its supernova remnant population inefficient cosmic ray accelerators. {\it Prima facie}, this is not unreasonable: the very large {\it volumetric-average} gas density of the region would be expected to imply strong ionisation cooling of low-energy cosmic rays, perhaps preventing their acceleration to higher energies.
Our detailed modelling of the region (\cite{Crocker2011b}), however, shows that this explanation also cannot work: GC SNRs are {\it at least} typically `efficient' as cosmic ray accelerators losing $\gtrsim 10$\% of their total $10^{51}$ erg mechanical energy into freshly-accelerated non-thermal particles.
\item Finally, it may be that the GC is not a calorimeter to either cosmic ray electrons or protons. 
\end{enumerate}

Given the exclusion of the first two possibilities, it seems that the third explanation given above must hold.
But here an immediate challenge is encountered: the diffuse, non-thermal radiation described above is very hard, consistent with $dN/dE \propto E^{-2.3}$ type population of both emitting electrons and protons.
This is quite different to the situation generally encountered in the Galactic disk (and directly measured at Earth): the cosmic ray spectrum is $\sim  E^{-2.7}$, significantly steeper than the expectation for the  distribution of cosmic rays injected into the ISM following diffusive (first order) Fermi acceleration at astrophysical shocks.
So some process acts to steepen the Galaxy's steady state cosmic ray population and -- has been known for decades -- a natural explanation of this steepening is that it is due to the energy-dependence of the confining effect of the Galaxy's magnetic field.
Cosmic rays scatter on magnetic field inhomogeneities, effectively diffusing through the Galaxy's volume but with a diffusion coefficient that (given the growth of a particle's gyroradius with momentum) grows with energy: high-energy particles are kept for a shorter time than low energy ones.
This cannot be happening in the GC: the hard spectrum of the in situ particle population is completely consistent with the expectation for the injection distribution.
Thus, if some process is acting to transport particles away -- as apparently required on the basis of the evidence described above -- this process must act without prejudice as to particle energy.

This requirement is naturally met by a large-scale outflow or wind of a few hundred km/s.
In fact, there is ample observational evidence for such an outflow (reviewed in \cite{Crocker2011b}) and observations of the star-forming external galaxies would tell us to expect one (e.g., \cite{Strickland2009}).
Indeed, scaling the results of \cite{Martin2005} according to the GC's estimated SFR areal density, we find that the expectation afforded by observations of external galaxies is that the GC should drive an outflow with a speed of $\sim$400 km/s.

Particularly compelling evidence comes from the radio continuum observations of \cite{Law2010} who traces emission along two radio continuum spurs north of Sgr A and Sgr C showing that these spurs form part of a larger, coherent structure (the GC Lobe) but, more relevantly, the emission from these structures exhibits a non-thermal spectrum that steepens with Galactic latitude.
This is a clear signal of the synchrotron ageing of an electron population as it carried away from its injection site in the plane.

Another interesting piece of evidence is the (apparent) existence of a `very-hot', diffuse plasma through the region as traced by X-ray emission (\cite{Koyama1989,Muno2004}).
This plasma -- with a temperature of $\sim 8$ keV -- would be substantially hotter than anything generally encountered in Galactic plane SNRs.
Whether it exists or is, in fact, an illusion created by the effect of many of faint, point-like sources (\cite{Revnivtsev2009}), remains controversial, however.
But it is certainly true that such hot, diffuse plasmas are detected in the star-forming nuclei of star-forming galaxies and are, in fact, required to provide (a large fraction of) the mid-plane pressure required to launch the outflows from these systems (e.g., \cite{Strickland2009}).

So it seems that there is a star-formation driven outflow out the GC.
With the understanding that this wind exists, at the back-of-the-envelope level we can employ the contrast between the observed and (calorimetrically) predicted luminosities of the GC in RC and $\gamma$-ray wavebands to tell us something about the systems' environmental parameters.
The basic considerations here are that the relevant radiative energy loss timescale must be slow enough with respect to the outflow transport time that we only detect (radiatively) a small fraction of the power injected into both cosmic ray protons and electrons (\cite{Crocker2010b}).
This means that if one of the parameters governing radiative losses -- the magnetic field for synchrotron losses of the electrons, the gas density for hadronic losses of the protons -- is `dialed-up', there must be commensurate rise in the outflow speed.

But the outflow speed cannot rise without bound: the kinetic and thermal power of the outflow can do no more than saturate that total mechanical power delivered by the region's supernovae (and, sub dominantly, its stellar winds).
On the other hand -- and speaking more loosely -- we have an expectation from observation of external star-burst nuclei that the thermalization efficiency of the outflow is not likely to be much below $\sim 10$\% (i.e., at least 10\% of the mechanical power delivered by supernovae ends up heating or moving the ISM: e.g., \cite{Strickland2009}).

These considerations  imply that the mean magnetic field and gas density the non-thermal particles encounter as they escape from the GC system fall in the rough ranges 100-300 $\mu$G and 5-20 cm$^{-3}$, respectively.
This first determination confirms that the ISM magnetic field is very strong as previously indicated.
The second is very interesting as it implies the cosmic rays `see' a gas density less than the volumetric average in the region which is $\sim$ 100 cm$^{-3}$, dominated by the very high density ($\gtrsim 10^4$ cm$^{-3}$) but small filling factor cores of the region's giant molecular cloud complexes.
Thus cosmic rays seems to be somehow excluded from the densest parts of the molecular clouds.
In fact, this is likely not so mysterious: the existence of the general outflow means that the particles only remain in the system for a limited time, a time presumably too short for them to either diffuse or be convected into the hearts of the highly turbulent molecular gas distributions.
This would make an interesting contrast between the GC and the situation recently claimed (\cite{Papadopoulos2010}) for star-burst systems where cosmic rays have been alleged to have a crucial role in changing the chemistry  in the densest molecular core regions, thus crucially altering the conditions for star formation.

\section{Connection to the Fermi Bubbles}

Given that most cosmic rays leave the system before they lose much energy, we can infer that the GC launches about $\sim 10^{39}$ erg/s into non-thermal particles into the Galactic bulge on an outflow.
Independent considerations suggest that it has been sustaining this activity for at least a few Gyr.
What is the implication of this?

The `Fermi Bubbles' constitute one of the most interesting recent discoveries (\cite{Su2010}) in high-energy astronomy: these are enormous structures, discovered in $\sim$ GeV $\gamma$-ray data from the $Fermi$-LAT, that extend $\sim$10 kpc north and south from the Galactic plane above the Galactic centre.
They are characterised by an unusually-hard spectrum, $dF_\gamma/dE_\gamma \propto E_\gamma^{-2.1}$ and have a total luminosity $4 \times 10^{37}$ erg/s.
Most researchers have considered the general idea that the $\gamma$-ray emission from these structures arises from the inverse-Compton (IC) emission from a (mysterious) population of cosmic ray electrons.
Given that the spectrum of the Bubbles displays no obvious variation with Galactic latitude, however, it is necessary that the photon background being up-scattered by this putative electron population is the CMB.
This, in turn, implies that the electrons have an energy scale $\sim$TeV and consequently short IC loss times, $\sim 10^6$ years.
So, given the vast extension of the Bubbles, these electrons either have to be delivered very quickly -- presumably on an AGN-outflow originating at Sgr A* (\cite{Guo2011}) --
or accelerated in-situ by first (\cite{Cheng2011}) or second-order (\cite{Mertsch2011}) Fermi acceleration processes.

We have recently considered an alternative explanation that -- because of the long loss times on the low-density plasma of the Bubbles -- escapes the difficulties facing any leptonic mechanism, namely, that the Bubbles' $\gamma$-ray emission arises from the hadronic collisions of a population of cosmic ray {\it protons} (and heavier ions) populating their interiors (\cite{Crocker2011a}).
This explanation requires i) (given adiabatic and ionisation energy losses) a total cosmic ray hadron power $\sim 10^{39}$ erg/s that ii) (essentially because of the same long loss time referred to above) has been injected quasi-continuously into the Bubbles for a timescale of $\gtrsim 8$ Gyr.
Note that exactly these requirements are matched by the GC CR outflow that we identified above {\it on the basis of completely independent considerations} to do with observations at radio continuum and TeV $\gamma$-ray wavebands of the inner $\sim 200$ pc of the Galaxy.
This putative solution fits nicely from a number of other perspectives: 
\begin{enumerate}

\item The hard-spectrum of the emission is also explained: by construction, the cosmic rays injected into the Bubbles are trapped so there is no energy-dependent loss process acting to modify the in-situ, steady state distribution away from the injection spectrum and the daughter $\gamma$-rays will trace this hard, parent proton distribution.

\item On the other hand, $\pi^0$-decay kinematics enforces a down-turn below $\sim$GeV on a spectral energy distribution plot of the emitted $\gamma$-radiation; such a downturn in robustly detected, at least qualitatively, in the Bubbles' spectra (\cite{Su2010}).

\item The total enthalpy of the Bubbles can be calculated to be $\sim 10^{57}$ erg -- this can be supplied by the GC outflow over the same long, multi-Gyr timescales required for the hadronic $\gamma$-ray scenario.

\item The total plasma mass of the Bubbles is $\sim 10^8 \msun$ (\cite{Su2010}) -- this mass can also be explained given the rate of mass flux in the GC outflow and assuming the same long timescales.

\item Dynamically, the Bubbles end up being slightly over-pressured but slightly under-dense with respect to the surrounding halo plasma, with internal energy density supplied approximately equally by cosmic rays and their interior hot plasma. They can be expected to rise slowly under buoyancy. 

\item The hadronic scenario naturally predicts concomitant secondary electron production within the Bubbles; these secondaries would synchrotron-radiate on the Bubble's magnetic field thereby explaining the coincident (at lower Galactic latitude) `WMAP haze' detected (\cite{Finkbeiner2004,Dobler2008}) at microwave frequencies. 

\end{enumerate}

Of course, all this requires that the Fermi Bubbles are very old structures -- almost as old as the Galaxy -- and that they can trap TeV cosmic rays for multi-Gyr timescales.
In fact,  our scenario implies that they would be calorimeters for GC activity over the history of the Milky Way.
This is an interesting prospect indeed.

\section{Conclusions}

\begin{enumerate}
\item Our modelling shows that the GC environment is an analogue to that found in star-bursts to the extent that we can show that the energy-density in both thermal and non-thermal ISM components is $\sim$2 orders of magnitude higher than typically encountered in the Galactic disk.
\item Another similarity to star-burst nuclei is the driving of a powerful outflow from the system; in the GC case, at least, this outflow does not reach escape velocity, however.
\item On the other hand, star-formation in the region seems to have been sustained at more-or-less the current rate for many billions of years: most star-formation that has taken place in the system is not really `bursty', though there are, of course, stochastic variations in the star-formation rate when one looks with sufficiently fine grain.
\item We find that cosmic rays do not penetrate into the cores of the giant molecular clouds. This is presumably because their short residence time in the region does not permit this. A contrasting situation where cosmic rays crucially modify ISM chemistry in star-forming regions has recently been claimed for star-bursts (\cite{Papadopoulos2010}).
\item On the other hand, ionising collisions experienced by the cosmic rays may explaining the anomalously hot and high ionisation state inferred for the (low density)  envelope molecular phase identified in $H_3^+$ absorption studies (\cite{Goto2008}).
\end{enumerate}

Finally we remark that all the high-energy activity identified above can be explained as a result, ultimately, of the power injected through the process of (massive) star-formation; the super-massive black hole is not {\it required} to have much influx beyond the inner few pc.
A final speculation, in fact, is that a significant consequence of the sustained, star-forming activity of the GC is to {\it prevent} significant activity of the black hole.

\section{Acknowledgements}
I gratefully acknowledge the contribution of my collaborators -- David Jones, Felix Aharonian, Casey Law, Fulvio Melia, Juergen Ott, and Tomo Oka
 -- to the research presented here and thank the organisers for the invitation to speak at SED2011.

\end{document}